\documentclass[12pt]{article}

\usepackage{amsmath,amsfonts,amssymb}

\def\ttau{\tilde{\tau}}

\def\tn{\tilde{n}}

\def\bA{\mathbf{A}}

\def\tx{\tilde{x}}

\def\be{\begin{equation}}

\def\hi{\hat{i}}
\def\hj{\hat{j}}
\def\ee{\end{equation}}

\def\bea{\begin{eqnarray}}

\def\eea{\end{eqnarray}}

\def\tmH{\tilde{\mH}}

\def\bY{\mathbf{Y}}

\def\mH{\mathcal{H}}

\def\tp{\tilde{p}}

\def\tpi{\tilde{\pi}}

\def\halpha{\hat{\alpha}}

\newcommand{\mG}{\mathcal{G}}

\newcommand{\hk}{\hat{k}}

\def\mV{\mathcal{V}}

\def \bAi{\left(\mathbf{A}^{-1}\right)}

\def \bA{\mathbf{A}}
\def\ta{\tilde{a}}

\def\mH{\mathcal{H}}

\newcommand{\ba}{\mathbf{a}}

\newcommand{\mL}{\mathcal{L}}

\def\tba{\tilde{\mathbf{a}}}

\def\pb #1{\left\{#1\right\}}

\begin{document}

	\begin{titlepage}

		\vskip 0.4 cm
		
		\begin{center}
			{\Large{ \bf Canonical Description of the Tachyon Kink in General Background
					and Hamiltonian for Dp-brane-anti-Dp-brane system}}
			
			\vspace{1em}  
			
			\vspace{1em} J. Kluso\v{n} 			
			\footnote{Email addresses:
				klu@physics.muni.cz (J.
				Kluso\v{n}) }\\
			\vspace{1em}
			\textit{Department of Theoretical Physics and
				Astrophysics, Faculty of Science,\\
				Masaryk University, Kotl\'a\v{r}sk\'a 2, 611 37, Brno, Czech Republic}
			
			\vskip 0.8cm
			
			%
			%
			%
			%
			%
			%
			
			\vskip 0.8cm
			
		\end{center}

		\begin{abstract}
In this short note we present analysis of the tachyon kink solution in the canonical
formalism. We also find Hamiltonian for simplified system of Dp-brane and anti-Dp-brane.

		\end{abstract}
		
		\bigskip
		
	\end{titlepage}
	
	\newpage

\section{Introduction and Summary}\label{first}
Unstable objects are very important parts of  string theory. Their conformal
field theory description and their significance was discovered in remarkable series of papers
by A. Sen \cite{Sen:1999mh,Sen:1999mg,Sen:1998tt}, for review and extensive list of references, see
\cite{Sen:2004nf}. They 
also led to renewed interest in the advanced topics of string theory, as for
example open string field theory
\cite{Sen:1999xm,Sen:1999nx}, background independent string theory \cite{Witten:1992qy,Kraus:2000nj,Takayanagi:2000rz}
  and so on
\footnote{For review, see for example \cite{Taylor:2003gn,Fuchs:2008cc}.}.

It is remarkable that many phenomena related to unstable D-branes can be captured
by low energy effective action despite of the fact that the mass of the tachyon 
is of the string scale
\cite{Sen:1999md,Garousi:2000tr,Bergshoeff:2000dq,Kluson:2000iy}. For example, it was shown that the spatial dependent tachyon
condensation leads to lower dimensional Dp-brane with agreement with the conformal
field theory and string field theory description
\cite{Sen:2003tm,Kluson:2005fj,Kluson:2005hd}. It was also shown that at the tachyon
vacuum unstable Dp-brane disappears and the dynamics reduces to the gas of fundamental
strings 
\cite{Sen:2000kd,Gibbons:2000hf,Kwon:2003qn}. Moreover, it was also shown that tachyon condensation in case of D-brane anti-D-brane system leads to emergence of codimension two stable D-brane again with agreement with conformal field theory description \cite{Sen:2003tm}.

The goal of this paper is to continue the study of the tachyon condensation on the world-volume of Dp-brane in  general background when we focus on the description of this problem 
using canonical formalism. In more details, we find Hamiltonian for covariant form of unstable D-brane when we show that it is given as sum of $p+1$ first class constraints
where $p+1$ is number of world-volume dimensions. Then we consider tachyon kink solution 
with the same dependence as was given in 
\cite{Sen:2003tm}, see also \cite{Kluson:2005fj}. We also specify that all remaining world-volume fields depend on remaining $p-$ coordinates. Due to the fact that 
we study this problem in the canonical formalism we also propose an ansatz for conjugate momenta that have more complicated form than conjugate coordinates. Then we insert this proposed ansatz to the canonical equations of motion and we show that these equations of motion are obeyed on conditions when the fluctuation fields obey the equations of motion that arise from the action for stable D(p-1)-brane. This result shows that the tachyon kink solution reproduces lower dimensional D(p-1)-brane even in the canonical formalism. We also discover the presence of the electric flux that is transverse to the kink and that has meaning of the fundamental string transverse to stable D(p-1)-brane. However this flux is localized in the transverse region of the size of order $a^{-1}$ where $a$ i the parameter that defines tachyon kink solution and that goes to $\infty$ in the end. As a result this electric flux goes to zero in the limit $a\rightarrow \infty$ with agreement with the observation given in \cite{Sen:2003bc}. 

As the second goal of this paper we focus on the canonical analysis of Dp-brane anti-Dp-brane action. The first proposal of the action for this system was given in 
\cite{Sen:2003tm}. Then an alternative derivation of this action was presented in 
\cite{Garousi:2004rd,Garousi:2007fn} where this action was derived by specific 
projection from non-abelian action for two non-BPS Dp-branes. However this proposed action was discussed in \cite{Israel:2011ut} where it was shown that it does not reproduce solution of the boundary field theory that describes this system. As a result
this system is not well understood yet. Despite of this fact we try to find canonical
form of the original form of the action \cite{Sen:2003tm}. However even in this case
we are not able to find corresponding Hamiltonian in the full generality simply from the fact that it is not possible to find exact form of the dependence of  tachyon on canonical variables. This problem is a consequence of the fact that the effective
action for this coupled system is the sum of two square root terms. For that reason 
we restrict ourselves to the case of the zero gauge fields on both worldvolumes of branes and also to the case when their separation is small. Then we will be able to find
closed form of the Hamiltonian for this system and discuss its properties. In particular, 
we analyse firstly the case when brane and anti-brane coincide so that the tachyon is pure time dependent and corresponds to rolling tachyon solution. The second solution 
corresponds to the situation when the tachyon is in its vacuum state and we study dynamics of relative coordinate when we show that it linearly depends on time which however means that it quickly breaks our approximation of the small separation of brane and anti-brane. 

Let us outline our results and suggest possible extension of this work. We found canonical action for non-BPS Dp-brane. Then we studied tachyon kink solution in the canonical formalism and we showed that this solution corresponds to stable D(p-1)-brane. Then we analysed action for Dp-brane anti-Dp-brane system and we found Hamiltonian for its simplified form. We mean that the detailed study of D-brane anti-Dbrane system should be goal of the future project. In particular, we would like to 
see whether it would be possible to find covariant form of the action suggested 
in \cite{Israel:2011ut} and how it is related to the actions proposed in 
\cite{Sen:2003tm} and  
\cite{Garousi:2004rd,Garousi:2007fn}. This problem is currently under study.

The structure of this paper is as follows. In the next section (\ref{second}) we 
find canonical form of the action for non-BPS Dp-brane and study tachyon kink solution. Then in section (\ref{third}) we analyse Dp-brane anti-Dp-brane system in the canonical formalism.

\section{Tachyon Kink on non-BPS Dp-Brane at Canonical Formalism}\label{second}
We start with the action for non-BPS Dp-brane that has the form
\cite{Sen:1999md,Bergshoeff:2000dq,Kluson:2000iy}
\begin{equation}\label{Sact}
S=-\ttau_p \int d^{p+1}e^{-\phi}\mV(T)\sqrt{-\det (g_{\alpha\beta}+l_s^2 F_{\alpha\beta}+\partial_\alpha T\partial_\beta T)}
\end{equation}
where 
\begin{equation}
g_{\alpha\beta}=g_{\mu\nu}\partial_\alpha x^\mu\partial_\beta x^\nu
\ , \quad F_{\alpha\beta}=\partial_\alpha A_\beta-\partial_\beta A_\alpha \ . 
\end{equation}
Further, $T$ is the tachyonic field defined on the world-volume of non-BPS Dp-brane,
$\mV(T)$ is its potential when we presume that $\mV$ is even function with three extremes, where $T_0 = 0$
is unstable maximum with $\mV (T_0)= 0$ while $\pm T_{min}$ are global minima of the potential with $\mV (\pm T_{min}) = 0$. Further, $x^\mu(\xi^\alpha),\mu,\nu=0,1,\dots,9$ are scalar fields on the world-volume of non-BPS Dp-brane that describe embedding of Dp-brane into target space-time and $A_\alpha,\alpha=0,1,\dots,p$ are gauge fields living
on the world-volume of non-BPS Dp-brane. Note that the $p+1$-dimensional world-volume is
labelled with coordinates $\xi^\alpha$. Finally $g_{\mu\nu}(x)$ is target space metric and $\phi$ is dilaton field that generally depends on $x$. For simplicity we will presume that $\phi$ is constant equal to $\phi_0$. Note also that $\ttau_p$ is non-BPS Dp-brane tension proportional to $l_s^{p+1}$ where $l_s$ is string length. 

We would like to study tachyon kink solution in the canonical formalism so that we should firstly determine corresponding Hamiltonian. From (\ref{Sact}) we find following conjugate momenta
\begin{eqnarray}\label{defmom}
& &p_\mu=\frac{\partial S}{\partial (\partial_0 x^\mu)}=
-\ttau_p \mV e^{-\phi_0}g_{\mu\nu}\partial_\alpha x^\nu \bAi_S^{\alpha 0}
\sqrt{-\det \bA} \ , \nonumber \\
& &\pi^\alpha=\frac{\partial S}{\partial(\partial_0 A_\alpha)}=
-\ttau_p \mV e^{-\phi_0}\bAi^{\alpha 0}_A\sqrt{-\det \bA} \ , \nonumber \\
& &p_T=\frac{\partial S}{\partial (\partial_0 T)}=
-\ttau_p \mV e^{-\phi_0}\partial_\alpha T\bAi^{\alpha 0}_S\sqrt{-\det \bA} \ , 
\nonumber \\
\end{eqnarray}
where 
\begin{equation}
\bA_{\alpha\beta}=g_{\alpha\beta}+\partial_\alpha T\partial_\beta T
+l_s^2 F_{\alpha\beta} \ , 
\end{equation}
and where $\bAi^{\alpha\beta}$ is matrix inverse to $\bA_{\alpha\beta}$ so that
$\bA_{\alpha\beta}\bAi^{\beta\gamma}=\delta_\alpha^\gamma$ and we also defined
\begin{equation}
\bAi^{\alpha\beta}_S=\frac{1}{2}(\bAi^{\alpha\beta}+\bAi^{\beta\alpha}) \ , \quad 
\bAi^{\alpha\beta}_A=\frac{1}{2}(\bAi^{\alpha\beta}-\bAi^{\beta\alpha}) \ .
\end{equation}
Note also that the definition of the canonical momenta $\pi^\alpha$ implies following primary constraint 
\begin{equation}
\pi^0\approx 0 \ . 
\end{equation} 
Using definition of the canonical momenta given in (\ref{defmom})
we easily find that the bare Hamiltonian density is equal to
\begin{equation}
\mH_B=p_T\partial_0 T+\pi^\alpha \partial_0 A_\alpha+p_\mu\partial_0 x^\mu-\mL=\pi^i
\partial_i A_0
\end{equation}
while this theory possesses $p+1$ primary constraints
\begin{eqnarray}
&&\mH_i=p_\mu\partial_i x^\mu+p_T\partial_iT+F_{ij}\pi^j \ , \quad 
\nonumber \\
&&\mH_\tau=p_\mu g^{\mu\nu}p_\nu+p_T^2+\pi^i(g_{ij}+\partial_i T\partial_j T)\pi^j+
\ttau_p^2\mV^2 e^{-2\phi_0}\det (g_{ij}+l_s^2 F_{ij}+\partial_i T\partial_j T )
\ . \nonumber \\
\end{eqnarray}
Then we can also write the action (\ref{Sact}) in the canonical form
\begin{eqnarray}\label{actcan}
& &S=\int d^{p+1}\xi (p_\mu\partial_0 x^\mu+\pi^i\partial_0 A_i+p_T\partial_0T-\nonumber \\
& &-N(p_\mu g^{\mu\nu}p_\nu+p_T^2+\pi^i(g_{ij}+\partial_i T\partial_j T)\pi^j+\ttau_p^2\mV^2 e^{-2\phi_0}\det (g_{ij}+l_s^2 F_{ij}+\partial_i T\partial_j T))-
\nonumber \\
& &-N^i(p_\mu\partial_i x^\mu+F_{ij}\pi^j+p_T\partial_iT)-\pi^i\partial_i A_0) \ , \nonumber \\
\end{eqnarray}
where $N$ and $N^i$ are Lagrange multipliers corresponding to the constraints $\mH_i,\mH_\tau$. Then performing variation of the action (\ref{actcan})
we determine corresponding equations of motion 
\begin{eqnarray}
& &\partial_0 x^\mu=2Ng^{\mu\nu}p_\nu+N^i\partial_i x^\mu \ , \nonumber \\
& & \partial_0 A_i=
2N \bA_{ij}^S\pi^j+N^jF_{ji}+\partial_i A_0 \ , \nonumber \\
& &\partial_0 p_\mu=
2\partial_i[N\pi^i g_{\mu\nu}\partial_j x^\nu\pi^j]-
N\pi^i\partial_i x^\rho \partial_\mu g_{\rho\sigma}
\partial_j x^\sigma \pi^j+\nonumber \\
& &+2\tau_p^2\partial_i[N e^{-2\phi_0} \mV^2h_{\mu\nu}\partial_j x^\nu
\bAi^{ji}_S\det \bA] +\partial_i[N^i p_\mu] \ , \nonumber \\
& &\partial_0 T=2Np_T+N^i\partial_i T \ , \nonumber \\
& &\partial_0 p_T=2\partial_i[N\pi^i \partial_j T\pi^j]
-2\tau_p^2 Ne^{-2\phi_0}\mV\frac{d\mV}{dT}\det \bA+\nonumber \\
& &+2\tau_p^2\partial_i[N e^{-2\phi_0}\mV^2 \partial_j T\bAi^{ji}_S
\det \bA]+\partial_i[N^i p_T] \ , \nonumber \\
& &\partial_0 \pi^i=2\tau_p^2\partial_j[N e^{-2\phi_0}\mV^2\bAi^{ij}_A
\det \bA]+\partial_k[N^k\pi^i]-\partial_k[N^i\pi^k] \ .
\nonumber \\
\end{eqnarray} 
Further, variation of the action (\ref{actcan}) with respect to $N$ and $N^i$ where
we treat them as independent variables imply constraints $\mH_i\approx 0 \ , \mH_\tau\approx 0$. In the same way variation of the action (\ref{actcan}) with respect to $A_0$ implies following Gauss law constraint
\begin{equation}
\mG\equiv \partial_i 
\pi^i\approx 0 \ . 
\end{equation}
Now we are ready to proceed to the search of the tachyon kink solution in the
canonical formalism. Analysis performed in the Lagrangian formalism shows that 
this kink solution should correspond to stable D(p-1)-brane
\cite{Sen:2003tm}, see also \cite{Kluson:2005fj,Kluson:2005hd}.  Following the same logic we will 
presume that the tachyon field has following dependence of the world-volume coordinates 
\begin{equation}\label{ansT}
T(\xi)=f(a(y-t(\xi^{\halpha}))) \ , 
\end{equation}
where $y\equiv \xi^p$ and where $\xi^{\halpha} \ , \halpha=0,1,\dots,p-1$. Then as in 
\cite{Sen:2003tm}
 we presume that $f(u)$ is arbitrary function of $u$ with the property that 
$f'(u)>0,\forall u$ and $f(\pm \infty)=\pm\infty$. Finally $a$ is free parameter that
we can send to $\infty$ in the end. 

Let us  further presume that $A_\alpha$ and $x^\mu$ do not depend on $y$ so that
\begin{equation}\label{ansx}
x^\mu(\xi^\alpha)=\tx^\mu(x^{\halpha}) \ , \quad 
A_{\halpha}(\xi^\alpha)=\ta_{\halpha}(\xi^{\halpha}) \ , \hi=1,\dots,p-1 \ 
\end{equation}
and without lost of generality we set $A_y=0$.
The ansatz for conjugate momenta has more complicated form
\begin{equation}\label{ansp}
p_T=af'\mV\tp_T \ , \quad 
p_\mu=\frac{\tau_p}{T_{p-1}}f'\mV\tp_\mu \  ,\quad \pi^{\halpha}=\frac{\tau_p}{T_{p-1}}af'\mV\tpi^{\halpha} \ , \quad 
\end{equation}
where $T_{p-1}$ is tension of stable D(p-1)-brane and the factor
$\frac{\ttau_p}{T_{p-1}}$ was chosen through dimensional reason 
since $p_\mu$ is $p+1$-dimensional density while $\tp_\mu$ is $p-$dimensional density. Further,  $\pi^y$ will be determined bellow. Finally we presume following ansatz
for $N$ and $N^i$
\begin{equation}
N=\frac{T_{p-1}}{\tau_p}\frac{1}{af'\mV}\tn(\xi^{\halpha}) \ , \quad 
N^{\hi}=\tn^{\hi}(\xi^{\halpha}) \ 
\end{equation}
while $N^y$ will be determined by equations of motion.

With this ansatz we find that the matrix $\bA_{ij}$ has the form
\begin{eqnarray}
\bA_{ij}=\left(\begin{array}{cc} \
\ba_{\hi \hj}+a^2f'^2\partial_{\hi}t\partial_{\hj}t& -a^2f'^2\partial_{\hi}t \\
-a^2f'^2 \partial_{\hj}t & a^2 f'^2 \\ 
\end{array}
\right) \ , \quad  \det \bA_{ij}=a^2 f'^2\det \ba_{\hi\hj} \ , \nonumber \\
\end{eqnarray}
where 
\begin{equation}
\ba_{\hi\hj}=g_{\mu\nu}\partial_{\hi}\tx^\mu\partial_{\hj}\tx^\nu+
l_s^2 f_{\hi\hj} \ , \quad  f_{\hi\hj}=\partial_{\hi}\ta_{\hj}-\partial_{\hj}\ta_{\hi} \ . 
\end{equation}
For further purposes we determine matrix inverse to $\bA_{ij}$ that has the form
\begin{equation}
\bA^{ij}=\left(\begin{array}{cc}
 \tba^{\hi\hj}
 & \tba^{\hi\hk}\partial_{\hk}t  \\
\partial_{\hk}t \tba^{\hk\hj} & \frac{1}{a^2f'^2}+\partial_{\hi}t \tba^{\hi\hj}
\partial_{\hj}t \\
\end{array}\right) \ , 
\end{equation}
where $\tba^{\hi\hj}$ is matrix inverse to $\ba_{\hi\hj}$ so that
\begin{equation}
\ba_{\hi\hj}\tba^{\hj\hk}=\delta_{\hi}^{\hj} \ . 
\end{equation}
Let u start with the equation of motion for $T$ that for (\ref{ansT}),(\ref{ansx}) and 
(\ref{ansp}) takes the form
\begin{eqnarray}\label{eqTans}
-af'\partial_0 t=af'\mV\tp_T+N^y af'-\tn^{\hi}\partial_{\hi}t af' \ .
\nonumber \\
\end{eqnarray}
To proceed further note that the equation of motion for $N^y$ implies
\begin{equation}
p_T\partial_y T+\partial_y x^\mu p_\mu+F_{y\hi}\pi^{\hi}=0
\end{equation}
that, since $\partial_y T\neq 0$ and since $\partial_y x^\mu=F_{y\hi}=0$ implies 
that $p_T=0$. Note that $F_{y\hi}=\partial_{y}A_{\hi}-\partial_{\hi}A_y=0$ due to the fact that $A_{\hi}$ does not depend on $y$ and since $A_y=0$. Now since $p_T=0$ the equation (\ref{eqTans})  allows us to express $N^y$ as
\begin{equation}
N^y=\tn^{\hi}\partial_{\hi}t-\partial_0 t \ . 
\end{equation}
As the next step we  proceed to the equation of motion for $\pi^{\hi}$. Inserting (\ref{ansT}),(\ref{ansx}) and (\ref{ansp}) into right side of this equation we get 
\begin{eqnarray}\label{eqpiansa}
&&2\ttau_p^2\partial_j[N e^{-2\phi_0}\mV^2 \bAi^{\hi j}\det \bA]
+\partial_k [N^k\pi^y]-\partial_k[N^{\hi}\pi^k]=\nonumber \\
& &af'\ttau_p\mV[2 T_{p-1}\partial_{\hj}[e^{-2\phi_0}\tn \tba^{\hi\hj}_A\det\ba]
+
\frac{1}{T_{p-1}}\partial_{\hk}[\tn^{\hk}\tpi^{\hi}]-\frac{1}{T_{p-1}}\partial_{\hk}[\tn^{\hi}\tpi^{\hk}]]+\nonumber \\
& &+\frac{\ttau_p}{T_{p-1}}\partial_y[af'\mV]
(-\partial_{\hk}t \tn^{\hk}\tpi^{\hi}+\tn^{\hk}\partial_{\hk}t\tpi^{\hi}-
\partial_0 t\tpi^{\hi}+\partial_{\hk}t\tn^{\hi}\tpi^{\hk})-\tn^{\hi}\partial_y
\pi^y\nonumber \\
\end{eqnarray}
while on the left-side  we have
\begin{equation}\label{eqpiansb}
\partial_0 \pi^{\hi}=-\frac{\ttau_p}{T_{p-1}}\partial_0t\partial_y[af'\mV]\tpi^{\hi}+\frac{\ttau_p}{T_{p-1}}af'\mV\partial_0\tpi^{\hi}
\end{equation}
using the fact that
\begin{equation}
\partial_{\halpha}[af'\mV]=-\partial_y[af'\mV]\partial_{\halpha}t\ . 
\end{equation}
Then collecting (\ref{eqpiansa}) together with (\ref{eqpiansb}) we obtain
\begin{eqnarray}\label{colpii}
& &af\mV\partial_0\tpi^i=af'\mV\left(2T_{p-1}^2\partial_{\hj}[\tn e^{-2\phi_0}\tba^{\hi\hj}_A\det\ba]
+
\partial_{\hk}[\tn^{\hk}\tpi^{\hi}]-\partial_{\hk}[\tn^{\hi}\tpi^{\hk}]\right)+
\nonumber \\
& &+\partial_y[af'V]\partial_{\hk}t\tn^{\hi}\tpi^{\hk}-\frac{T_{p-1}}{\ttau_p}\tn^{\hi}\partial_y \pi^y
	\nonumber \\
\end{eqnarray}
We see that the expression on the second line vanishes when we presume that $\pi^y$ is equal to
\begin{equation}
	\pi^y=\frac{\ttau_p}{T_{p-1}}af'\mV\partial_{\hi}t\tpi^{\hi} \ . 
	\end{equation}
	Then (\ref{colpii}) has final form 
\begin{equation}
af'\mV(\partial_0\tpi^{\hi}-2T_{p-1}^2\partial_{\hj}[\tn e^{-2\phi_0} \tba^{\hi\hj}_A\det\ba]
-
\partial_{\hk}[\tn^{\hk}\tpi^{\hi}]+\partial_{\hk}[\tn^{\hi}\tpi^{\hk}])=0 \ .
\end{equation}
Now this expression is equal to zero in the limit $a\rightarrow \infty $ for $y\neq t(\xi^{\halpha})$. On the other hand we find that the expression $
af'\mV$ is finite for $y=t(\xi^{\halpha})$ and hence in order to obey equations of motion for $\pi^{\hi}$ at the core of the kink defined as $y^p=t(\xi^{\halpha})$  we should demand that the expressions in the bracket is zero. However they are precisely the  equations of motion for $\tpi^{\hi}$ that follow from an action for D(p-1)-brane. In more details, let us consider canonical form of the action for stable D(p-1)-brane
\begin{equation}\label{Sactstable}
S=-T_{p-1}\int d^p\xi
(\tp_\mu\partial_0\tx^\mu+\tpi^{\hi}\partial_0\ta_{\hi}-\tn \tmH-
\tn^{\hi}\tmH_{\hi}-\tpi^{\hi}\partial_{\hi} \ta_0) \ , 
\end{equation}
where
\begin{eqnarray}
& &\tmH=\tp_\mu g^{\mu\nu}\tp_\nu+\tpi^{\hi}g_{\hi\hj}\tpi^{\hj}+T_{p-1}^2e^{-2\phi_0}\det (g_{\hi\hj}+l_s^2 f_{\hi\hj}) \ , 
\nonumber \\
& &\tmH_{\hi}=\tp_\mu\partial_{\hi} \tx^\mu+f_{\hi\hj}\tpi^{\hj} \ , \quad g_{\hi\hj}=
g_{\mu\nu}\partial_{\hi}\tx^\mu\partial_{\hj}\tx^{\nu} \ , \quad 
f_{\hi\hj}=\partial_{\hi}\ta_{\hj}-\partial_{\hj}\ta_{\hi} \ . 
\end{eqnarray}
Then from the action (\ref{Sactstable}) we obtain following  equations of motion for $\tx^\mu$
\begin{equation}\label{eqmxDp}
\partial_0 \tx^\mu
=2\tn g^{\mu\nu}\tp_\nu+\tn^{\hi}\partial_{\hi}\tx^\mu \ 
\end{equation}
and for $\ta_{\hi}$
\begin{equation}\label{eqmaDp}
\partial_0 \ta_{\hi}=
2\tn g_{\hi\hj}\tpi^{\hj}+\tn^{\hj}f_{\hj\hi}+\partial_{\hi}\ta_0 \ , 
\end{equation}
together with equations of motion for canonically conjugate momenta $\tp_\mu$
\begin{eqnarray}\label{eqmpDp}
& &\partial_0 \tp_\mu=
2\partial_{\hi}[\tn\tpi^{\hi} g_{\mu\nu}\partial_{\hj} \tx^\nu\tpi^{\hj}]-
\tn\tpi^{\hi}\partial_{\hi} \tx^\rho \partial_\mu g_{\rho\sigma}
\partial_{\hj} \tx^\sigma \tpi^{\hj}+
\nonumber \\
&&+T_{p-1}^2 e^{-2\phi_0}
\partial_{\hi}\tx^\rho \partial_\mu g_{\rho\sigma}
\partial_{\hj}\tx^\nu \tba^{\hj\hi}\det \ba +\nonumber \\
& &+2T_{p-1}^2\partial_{\hi}[\tn e^{-2\phi_0} g_{\mu\nu}\partial_{\hj} \tx^\nu
\tba^{ji}_S\det \ba] +\partial_{\hi}[\tn^{\hi} p_\mu] \  \nonumber \\
\end{eqnarray}
and $\tpi^{\hi}$
\begin{eqnarray}\label{eqmpiDp}
\partial_0 \tpi^{\hi}=2T_{p-1}^2\partial_{\hj}[\tn e^{-2\phi}\tba^{\hi
\hj}_A
\det \ba]+\partial_{\hk}[\tn^{\hk}\tpi^{\hi}]-\partial_{\hk}[\tn^{\hi}\tpi^{\hk}] \ . 
\end{eqnarray}
Now we return back to non-BPS Dp-brane and
consider equation of motion for $\pi^y$ 
\begin{equation}
\partial_0\pi^y=2\ttau_p^2\partial_j[N e^{-2\phi_0}\mV^2\bAi^{yj}_A\det \bA]
+\partial_k[N^k\pi^y]-\partial_k[N^y\pi^k]  \ . 
\end{equation}
Inserting ansatz (\ref{ansT}),(\ref{ansx}) and
(\ref{ansp}) into it we obtain after some calculations following result 
\begin{eqnarray}\label{anspy}
af'\mV\partial_{\hi}t[\partial_0\tpi^{\hi}-T_{p-1}^2
\partial_{\hk}[
\tn e^{-2\phi_0}\tba_A^{\hi\hk}\det\ba]+\partial_{\hk}[\tn^{\hi}\tpi^{\hk}]-
\partial_{\hk}[\tn^{\hk}\tpi^{\hi}]]=0 \ , 
\nonumber \\
\end{eqnarray}
where we also used the fact that 
\begin{equation}
\bAi^{y\hi}_A
=\frac{1}{2}(\partial_{\hk}t\tba^{k\hi}-\tba^{\hi\hk}\partial_{\hk}t)
=\partial_{\hk}t\tba_A^{\hk\hi} \ . 
\end{equation}
It is clear that  (\ref{anspy}) is obeyed  when the equations
of motion (\ref{eqmpiDp}) hold. This is again nice consistency check. 

As the next step we proceed to the equation of motion for $p_T$. Note that
the left side is equal to zero since $p_T=0$ while the right side is equal to
\begin{eqnarray}\label{eqTright}
& &2\partial_i[N\pi^i \partial_j T\pi^j]
-\ttau_p^2 Ne^{-2\phi_0}\mV\frac{d\mV}{dT}\det \bA+\nonumber \\
&&+\ttau_p^2\partial_i[N e^{-2\phi_0}\mV^2 \partial_j T\bAi^{ji}_S
\det \bA]+\partial_i[N^i p_T]  \ . \nonumber \\
\end{eqnarray}
Let us  consider the  first term on the first line in (\ref{eqTright}). Inserting
(\ref{ansT}),(\ref{ansx}) and (\ref{ansp}) into it  we obtain that it identically vanishes since
\begin{eqnarray}
&&\partial_i[N\pi^i\partial_j T\pi^j]
=\frac{\ttau_p}{T_{p-1}}\left(\partial_y[a^2f'^2\mV]\tpi^{\hi}\partial_{\hi}t\tpi^{\hk}\partial_{\hk}t
-\partial_y[a^2f'^2\mV]\tpi^{\hi}\partial_{\hi}t\partial_{\hk}t\tpi^{\hk}\right.+\nonumber \\
& &+\left.+\partial_{\hi}[a^2f'^2\mV
\tpi^{\hi} \tpi^{\hk}\partial_{\hk}t]-\partial_{\hi}[a^2f'^2 \mV\tpi^{\hi}\partial_{\hj}t\tpi^{\hj}]\right)=0 \ . 
\nonumber \\
\end{eqnarray}
Let us now insert (\ref{ansT}),(\ref{ansx}) and (\ref{ansp}) into the first term on the second line in (\ref{eqTright}) 
and we obtain
\begin{eqnarray}
& &\partial_{i}[N e^{-2\phi_0}\mV^2\partial_j T\bAi^{ji}_S\det\bA]=\nonumber \\
& &=\ttau_p T_{p-1}\partial_y[\tn a^2f'^2e^{-2\phi_0}\mV\det\ba
(\bAi^{yy}-\partial_{\hj}t\bAi^{jy}_S)]+\nonumber \\
&&+\ttau_p T_{p-1}\partial_{\hi}[\tn a^2f'^2e^{-2\phi_0}\mV
\det\ba (\bAi^{y\hi}_S-\partial_{\hj}t\bAi^{\hj\hi}_S)]=\nonumber \\
& &=\ttau_p T_{p-1}\tn e^{-2\phi_0}af'\frac{d\mV}{dT}\det \ba  \nonumber \\
\end{eqnarray}
which exactly cancel the second term on the first line in (\ref{eqTright}). As a result we obtain 
that the equation of motion for $p_T$ is identically obeyed for the ansatz
(\ref{ansT}),(\ref{ansx}) and (\ref{ansp}). 

As the next step we consider equation of motion for $A_\alpha$
\begin{equation}
\partial_0 A_{\hi}=2N\bA_{\hi j}^S\pi^{j}+N^j F_{ji}+\partial_{\hi}A_0
\end{equation}
that for the ansatz (\ref{ansT}),(\ref{ansx}) and (\ref{ansp}) reduces into
\begin{equation}
af'\mV\left(\partial_0 \ta_{\hi}-\partial_{\hi}\ta_0-2\tn \tba_{\hi\hj}^S\tpi^{\hj}-\tn^{\hj}f_{\hj\hi}\right)=0
\end{equation}
and we see that this equation is obeyed at the core of the kink on condition when the
expression in the bracket vanishes. In fact, this expression is the equation of motion for
the gauge field $\ta_{\hi}$ given in  (\ref{eqmaDp}).

In the same way we can proceed with the equation of motion for $x^\mu$ and we obtain 
\begin{eqnarray}
af'\mV\left(\partial_0 \tx^\mu-2\tn g^{\mu\nu}\tp_\nu-\tn^{\hi}\partial_{\hi}\tx^\mu\right)=0 \nonumber \\
\end{eqnarray}
and we see that this equation of motion is obeyed at the core of the kink when the expression 
in the bracket vanishes. Again, this expression has the same form as the equation of motion for 
$\tx^\mu$ (\ref{eqmxDp}).

The situation is more involved in case of the equation of motion for $p_\mu$. Inserting
the ansatz (\ref{ansp})) to the left side of this equation we obtain
\begin{equation}\label{leftsideeqmp}
\partial_0 p_\mu=\frac{\ttau_p}{T_{p-1}}(-\partial_y (af'\mV)\partial_0 t\tp_\mu+
af'\mV\partial_0\tp_\mu
) \ . 
\end{equation}
On the other hand when we insert (\ref{ansT}),(\ref{ansx}) and (\ref{ansp}) to the right side of the equation of motion for $p_\mu$ we obtain
\begin{eqnarray}\label{rightsideeqmp}
& &2\partial_i[N\pi^i g_{\mu\nu}\partial_j x^\nu\pi^j]-
N\pi^i\partial_i x^\rho \partial_\mu g_{\rho\sigma}
\partial_j x^\sigma \pi^j
-\ttau_p^2 N e^{-2\phi_0}\partial_i x^\rho\partial_\mu g_{\rho\sigma}\partial_j x^\sigma\bAi^{ji}
\det \bA
+\nonumber \\
& &+2\ttau_p^2\partial_i[N e^{-2\phi_0}\mV^2 g_{\mu\nu}\partial_j x^\nu
\bAi^{ji}_S\det \bA] +\partial_i[N^i p_\mu]= \nonumber \\
& &2\frac{\ttau_p}{T_{p-1}}af'\mV
\partial_{\hi}[\tn e^{-2\phi_0} g_{\mu\nu}\partial_{\hj}\tx^\nu\tpi^{\hj}]-
\frac{\ttau_p}{T_{p-1}}af'V\tpi^{\hi}\partial_{\hi}\tx^\rho
\partial_\mu g_{\rho\sigma}\partial_{\hj}\tx^\sigma \tpi^{\hj}+\nonumber \\
&&+
2af'\mV\ttau_p T_{p-1}\partial_{\hi}[\tn e^{-2\phi_0} g_{\mu\nu}\partial_{\hj}\tx^\nu
\tba^{\hj\hi}_S\det\ba] 
-
\nonumber \\
&&-\frac{\ttau_p}{T_{p-1}}\partial_y(af'\mV)\partial_{\hi}t \tn^{\hi}\tp_\mu+
\frac{\ttau_p}{T_{p-1}}af'\mV \partial_{\hi}[\tn^{\hi}\tp_\mu]+
\frac{\ttau_p}{T_{p-1}}\partial_y(af'\mV)(\tn^{\hi}\partial_{\hi}t-\partial_0t)\tp_\mu \ ,  \nonumber \\
\end{eqnarray}
where we used 
\begin{eqnarray}
\partial_i[N\pi^ig_{\mu\nu}\partial_j x^\nu\pi^j]=
\frac{\ttau_p}{T_{p-1}}af'\mV
\partial_{\hi}[\tn g_{\mu\nu}\partial_{\hj}\tx^\nu\tpi^{\hj}]\nonumber \\
\end{eqnarray}
and also
\begin{eqnarray}
\ttau_p^2\partial_i[N g_{\mu\nu}\partial_j x^\nu \bAi^{ji}_S\det\bA]=
af'\mV\ttau_p T_{p-1}\partial_{\hi}[\tn g_{\mu\nu}\partial_{\hj}\tx^\nu
\tba^{\hj\hi}_S\det\ba] \ .
\nonumber \\
\end{eqnarray}
Taking (\ref{leftsideeqmp}) together with (\ref{rightsideeqmp}) we obtain 
\begin{eqnarray}
& &af'\mV\left(-\partial_0\tp_\mu+
\partial_{\hi}[\tn g_{\mu\nu}\partial_{\hj}\tx^\nu\tpi^{\hj}]-
\tpi^{\hi}\partial_{\hi}\tx^\rho
\partial_\mu g_{\rho\sigma}\partial_{\hj}\tx^\sigma \tpi^{\hj}+\right.\nonumber \\
& &+\left.T_{p-1}^2 e^{-2\phi_0}
\partial_{\hi}\tx^\rho \partial_\mu g_{\rho\sigma}
\partial_{\hj}\tx^\nu \tba^{\hj\hi}\det \ba+ 2 T^2_{p-1}\partial_{\hi}[\tn e^{-2\phi_0} g_{\mu\nu}\partial_{\hj}\tx^\nu
\tba^{\hj\hi}_S\det\ba] 
+ \partial_{\hi}[\tn^{\hi}\tp_\mu]\right)=0 
\nonumber \\
\end{eqnarray}
and we see that the equation of motion for $p_\mu$ is obeyed at the core of the kink
on condition when the fluctuation fields around the kink solution obey the equations
of motion (\ref{eqmpDp}).

In summary, we have shown that the tachyon kink solution describes D(p-1)-brane at the core of the kink when the kink is localized along $y-$direction at the point $t(\xi)$. Note that $t(\xi^\alpha)$ can be completely arbitrary which is manifestation of the diffeomorphism invariance of the world-volume theory. On the other hand non-trivial profile $t(\xi)$ rises a puzzle since it induces, according to the analysis presented above, non-zero electric field equal to
\begin{equation}
\pi^y=\frac{\ttau_p}{T_{p-1}}af'\mV\tpi^{\hi}\partial_{\hi}t \ .
\end{equation}
It is remarkable that when $t(\xi^{\halpha})$ is constant then the electric flux along
$y-$ direction is zero. In fact, this electric flux is localized around the core of the kink due to the presence of the factor $af'\mV$ which is in agreement with the result derived in \cite{Sen:2003bc}. 

\section{Canonical Analysis $Dp-\overline{D}p$-brane action}\label{third}
In this section we perform canonical analysis of the action for Dp-brane anti-Dp-brane
system at least in its simplified form.  

 The first proposal for the action that describes
this system was given by A. Sen in \cite{Sen:2003tm}. It was suggested there that
such an action has the form
\footnote{For related works, see
\cite{Hatefi:2017ags,Hatefi:2012cp,Israel:2011ut,Erkal:2009xq,Garousi:2007fn,
Garousi:2004rd}.}
\begin{equation}
S=-\int d^{p+1}\xi V(T,X_{(1)}-X_{(2)})
(\sqrt{-\det \bA_{(1)}}+\sqrt{-\det \bA_{(2)}}) \ ,
\end{equation}
where 
\begin{eqnarray}
& &\bA_{(i)\mu\nu}=\eta_{\mu\nu}+\partial_\mu X^{I}_{(i)}\partial_\nu X^{I}_{(i)}+
F_{\mu\nu}^{(i)}+\frac{1}{2}(D_\mu T)^*D_\nu T+
\frac{1}{2}(D_\nu T)^*D_\mu T \ , \nonumber \\
& &F_{\mu\nu}^{(i)}=\partial_\mu A_\nu^{(i)}-\partial_\nu A_\mu^{(i)} \ , \quad 
D_\mu T=(\partial_\mu -iA_{\mu}^{(1)}+iA_\mu^{(2)})T \ , \nonumber \\
\end{eqnarray}
where $(i)=(1),(2)$ where $(1)$ corresponds to the first Dp-brane while $(2)$ corresponds to the
second one. Further, A. Sen suggested that the  tachyon potential has the form 
\begin{equation}\label{potpp}
V(T,X_{(1)}-X_{(2)})=
T_{p+1}\left[1+\frac{1}{2}\sum_I\frac{(X_{(1)}^I-X_{(2)}^I)^2}{l_s^2})T^2+O(T^4)\right]
\end{equation}
It is important to stress that these Dp-brane actions were formulated in the
static gauge when 
\begin{equation}\label{statgauge}
	\xi^\alpha=X^{\alpha}_{(1)}=X^{\alpha}_{(2)}  \ . 
\end{equation}
Then it is natural to generalize this action to the manifestly diffeomorphism invariant
form when we write it as
\begin{eqnarray}
&&	\bA_{(i)\alpha\beta}=\partial_\alpha X^M_{(i)}\partial_\beta X^N_{(i)}\eta_{MN}+l_s^2F_{\alpha\beta}^{(i)}+
\frac{1}{2}(D_\alpha T)^*D_\beta T+
\frac{1}{2}(D_\beta T)^*D_\alpha T	
 	 \ , 
	\nonumber \\
&&	V(T,X_{(1)}-X_{(2)})=T_p
\left[1+\frac{1}{2}\sum_M \frac{(X^M_{(1)}-X^M_{(2)})^2}{l_s^2})T^2+O(T^4)\right] \ , 
	\nonumber \\
\end{eqnarray}
where $(X^M_{(1)}-X^M_{(2)})^2=(X^M_{(1)}-X^M_{(2)})\eta_{MN}(X^N_{(1)}-X^N_{(2)})$ that in the static gauge (\ref{statgauge}) the potential given above reduces into 
	(\ref{potpp}). 

It is very difficult to find canonical form of this action in the full generality due to the fact that both determinants $\det \bA_{(1)}$ and $\det \bA_{(2)}$ contain kinetic term for tachyon and we were not able to find Hamiltonian constraint. For that reason we restrict ourselves to simpler case when the gauge fields $A_\alpha^{(1)}$ and $A_{\alpha}^{(2)}$ vanish. 
We further presume that the  separation of D-brane anti-D-brane pair is small so that when we write $X^M_{(1)}=Y^M+\frac{l^M}{2} \ , X^M_{(2)}=Y^M-\frac{l^M}{2}$ we obtain
\begin{eqnarray}
\sqrt{-\det \bA_{1}}+\sqrt{-\det \bA_{(2)}}=\sqrt{-\det \bY}(2+\frac{1}{4}
\bY^{\alpha\beta}\partial_\alpha l^M\partial_\beta l_M) \ ,
\end{eqnarray}
where
\begin{equation}
\bY_{\alpha\beta}=g_{\alpha\beta}+\partial_\alpha \rho \partial_\beta \rho+\rho^2
\partial_\alpha \phi \partial_\beta \phi \ , 
\end{equation}
and where we introduced following   parametrization of 
$T$ 
\begin{equation}
T=\rho e^{i\phi} \ .
\end{equation}
Note that the potential $V$ is now function of $\rho$ only.
 
To proceed further we introduce common notation $Z^P=(Y^M,\rho,\phi)$ and matrix $H_{PQ}=
\mathrm{diag}(\eta_{MN},1,\rho^2)$ so that the matrix $\bY_{\alpha\beta}$ has
the form $\bY_{\alpha\beta}=H_{PQ}\partial_\alpha Z^P \partial_\beta Z^Q$.
%
In order to find canonical formulation of this theory we use ADM like decomposition of the matrix
$\bY_{\alpha\beta}$ in the form
\begin{equation}
	\bY_{\alpha\beta}=\left(\begin{array}{cc}
		-M^2+M_i m^{ij}M_j & M_j \\
		M_i & m_{ij} \\ \end{array}\right) \ , \quad \bY^{\alpha\beta}=
	\left(\begin{array}{cc}
-\frac{1}{M^2}	& \frac{M^j}{M^2} \\
\frac{M^i}{M^2} & m^{ij}-\frac{M^i M^j}{M^2} \\ \end{array}\right) \ , 
\end{equation}
where $m^{ij}$ is matrix inverse to $m_{ij}$ so that $m_{ij}m^{jk}=\delta_i^k$.
Let us compare this notation with the definition $\bY_{\alpha\beta}=\partial_\alpha Z^P H_{PQ}\partial_\beta Z^Q$ and we obtain 
\begin{eqnarray}\label{defMMi}
& &	m_{ij}=\partial_i Z^P H_{PQ}\partial_jZ^Q \ , \quad  M_i=\partial_i Z^P H_{PQ}\partial_0 Z^Q \ , 
	\nonumber \\
& &	M^2=
%
	 -\partial_0 Z^P V_{PR}\partial_0 Z^R \ , \quad 
	V_{PR}=H_{PR}-H_{PQ}\partial_i Z^Q m^{ij}\partial_j Z^S H_{SR} \ . \nonumber \\
		\end{eqnarray}
With this notation the action has the form
\begin{eqnarray}\label{SactM}
& &	S=-\int d^{p+1}\xi V M \sqrt{\det m_{ij}}
(2-\frac{1}{4}\nabla_m l^M \nabla_m l_M+\frac{1}{4}m^{ij}\partial_i l^M\partial_j l_M)	 \ , 
\nonumber \\
& &\nabla_m l=\frac{1}{M}(\partial_0 l-M^i \partial_i l)  \ . \nonumber \\
\end{eqnarray}
Now we  are ready to find conjugate momenta. From (\ref{SactM}) we obtain
momentum conjugate to $l^M$ in the form
\begin{equation}
p_M^l=\frac{\delta S}{\delta \partial_0 l^M}=
\frac{1}{2}V\sqrt{\det m_{ij}}\eta_{MN}\nabla_m l^N \ .
\end{equation}
In case of the fields $Z^P$ we should be more careful since $M$ and $M_i$ depends on $\partial_0 Z^P$. Explicitly,  from (\ref{defMMi}) we obtain  
\begin{equation}
\frac{\partial M}{\partial \partial_0 Z^P}=-\frac{V_{PQ}\partial_0 Z^Q}{N} \ , 
\quad \frac{\partial M_i}{\partial \partial_0 Z^P}=H_{PQ}\partial_i Z^Q \ .
\end{equation}
Then, after some calculations, we get
\begin{eqnarray}
& &\Pi_P\equiv (K_P+H_{PQ}\partial_i Z^Q m^{ij}\partial_j l^M p_M^l)=\nonumber \\
& &=V\frac{V_{PQ}\partial_0 Z^Q}{M}\sqrt{\det m_{ij}}
(2-\frac{1}{V^2 \det m_{ij}}p^l_Mp^l_N\eta^{MN}+
\frac{1}{4}m^{ij}\partial_il^M\partial_jl^N\eta_{MN})
+\nonumber \\
&&2\frac{V}{M}V_{PQ}\partial_0 Z^Q \frac{1}{V^2 \det m_{ij}}p_M^l p^l_N\eta^{MN}\ ,  \nonumber \\
\end{eqnarray}
where $K_P$ is momentum conjugate to $Z^P$. 
To proceed further we use the fact that 
\begin{equation}\label{iden}
\partial_i Z^P V_{PQ}=0 \ , \quad 
V_{PQ}H^{QR}V_{RT}=V_{RT} \ . 
\end{equation}
Using the first relation given above we obtain 
\begin{equation}
\partial_i Z^P K_P=-\partial_i l^M p_M^l
\end{equation}
that implies following $p-$primary constraints
\begin{equation}
\mH_i\equiv \partial_i Z^P K_P+\partial_i l^M p_M^l\approx 0 \ . 
\end{equation}
Further, the second relation in (\ref{iden}) implies
\begin{eqnarray}
\Pi_P H^{PQ}\Pi_Q
=-V^2\det m_{ij}[4+m^{ij}\partial_il^M \partial_j l_N]-4p_M^l p_N^l\eta^{MN} 
\nonumber \\
\end{eqnarray}
that implies an existence of Hamiltonian constraint
\begin{eqnarray}
	\mH_\tau=\Pi_P H^{PQ}\Pi_Q+4p^M_l p^N_l\eta_{MN}+V^2\det m_{ij}[4+m^{ij}\partial_i l^M\partial_j l_M]\approx 0
	\nonumber \\
	\end{eqnarray}
or explicitly
\begin{eqnarray}
\mH_\tau=p_M \eta^{MN}p_N+p_\rho^2+\frac{1}{\rho^2}p_\phi^2+4p_M^l p_N^l\eta^{MN}	
+
V^2\det m_{ij}[4+m^{ij}\partial_i l^M\partial_j l_M]\approx 0 \ , \nonumber \\
	\end{eqnarray}
where we used 	the fact that $\partial_i Z_P K^P=\mH_i-p_M^l\partial_i l^M$ and then we neglected terms of the fourth order in $p_M^l$ and $l_M^l$.
In summary, we get Hamiltonian as sum of $p+1$ primary constraints
\begin{equation}
H=\int d^p\xi \mH \ , \quad \mH=N\mH_\tau+N^i\mH_i \ . 
\end{equation}
 Using this Hamiltonian 
we can study dynamics of Dp-brane anti-Dp-brane system. 
The canonical equations of motion for $Y^M$ and $p_M$ have the form
\begin{eqnarray}
& &	\partial_0 Y^M=\pb{Y^M,H}=2N\eta^{MN}p_N+N^i\partial_i Y^M \ , \nonumber \\
&&	\partial_0 p_M=\pb{p_M,H}=2\partial_i[NV^2 \eta_{MN}\partial_j Y^N
	m^{ji}\det m_{mn}(4+m^{kl}\partial_k l^M\partial_l l_M)]-
	\nonumber \\
&&-2\partial_k[NV^2\det m_{mn}m^{ik}\eta_{MN}\partial_l Y^N
m^{lj}\partial_i l^K\partial_j l_K]+\partial_i (N^i p_M) \ . \nonumber \\	
	\end{eqnarray}
In case of $\phi$ and $p_\phi$ we get
\begin{eqnarray}
&&\partial_0 \phi=\pb{\phi,H}=2\frac{N}{\rho^2}p_\phi+N^i\partial_i\phi  \ , 
\nonumber \\
&&\partial_0 p_\phi=\pb{p_\phi,H}=\partial_i (N^i p_\phi)+
\nonumber \\
&&+2\partial_i[NV^2 \partial_j \phi
m^{ji}\det m_{mn}(4+m^{kl}\partial_k l^M\partial_l l_M)]-
\nonumber \\
&&-2\partial_k[NV^2\det m_{mn}m^{ik}\partial_l \phi
m^{lj}\partial_i l^K\partial_j l_K] \ . \nonumber \\
\end{eqnarray}
The situation is slightly more involved in case of $\rho$ and $p_\rho$ when we have
\begin{eqnarray}
& &\partial_0\rho=\pb{\rho,H}=2Np_\rho+N^i\partial_i \rho \ , \nonumber \\
&&\partial_0 p_\rho=\pb{p_\rho,H}=4\frac{N}{\rho^3}p_\phi^2-2V\frac{dV}{d\rho}\det m_{ij}
[4+m^{ij}\partial_i l^M \partial_j l_M]+
\nonumber \\
&&+2\partial_i[NV^2 \partial_j p_\rho
m^{ji}\det m_{mn}(4+m^{kl}\partial_k l^M\partial_l l_M)]-
\nonumber \\
&&-2\partial_k[NV^2\det m_{mn}m^{ik}\partial_l \rho
m^{lj}\partial_i l^K\partial_j l_K]+\partial_i (N^i p_M) -\nonumber \\
&&-2NV^2\det m_{mn}\partial_i\phi\partial_j\phi m^{ji}(4+m^{kl}\partial_k l^M
\partial_l l_M)+\nonumber \\
&&+2NV^2\det m_{mn}m^{ik}\rho \partial_k\phi\partial_l\phi m^{lj}
\partial_i l^M\partial_j l_M +\partial_i (N^i p_\rho) \ . \nonumber \\
\end{eqnarray}
Finally the equations of motion for $l^M$ and $p_M^l$ have the form 
\begin{eqnarray}
&&\partial_\tau l^M=\pb{l^M,H}=
8N \eta^{MN}p_N^l+N^i\partial_i l^M \ , \nonumber \\
&&\partial_\tau p_M^l=
\pb{p_M^l,H}=-2V\frac{\partial V}{\partial l^M}\det m_{ij}
[4+m^{ij}\partial_il^M\partial_j l^M\eta_{MN}]+\nonumber \\
&&+2\partial_i[V^2\det m_{mn}m^{ij}\eta_{MN}\partial_j l^N]+\partial_i(N^i p_M^l)  \ . 
\nonumber \\
\end{eqnarray}
Let us now solve these equations. First of all we impose 
 static gauge 
\begin{equation}\label{Ystat}
	Y^\alpha=\xi^\alpha\ , \quad  l^\alpha=0 \ , \quad \alpha=0,1,\dots,p \ .
\end{equation}
Let us also presume that remaining fields depend on $\xi^0$ only. Then the spatial diffeomorphism
constraints imply
\begin{equation}
	\mH_i=p_i=0 \ , i=1,\dots,p  \ . 
\end{equation}
Further, (\ref{Ystat}) and the fact that $\rho$ and $\phi$ depend on $\xi^0$ only implies that $m_{ij}$ is $p\times p$ diagonal matrix
\begin{equation}
	m_{ij}=	\partial_i Y^M\eta_{MN}\partial_j Y^N=
	\delta_{ij} \ .
\end{equation}
Let us now consider equation of motion for $Y^0$ 
\begin{equation}
	\partial_0 Y^0=-2Np_0 
\end{equation}
that using (\ref{Ystat}) fixes $N$ to be equal to $N=-\frac{1}{2p_0}$. Further, inserting (\ref{Ystat}) into 
equation of motion for $Y^i$ and using $p_i=0$ we obtain that $N^i=0$.
Finally, equations of motion for $p_0$ implies
\begin{eqnarray}
	\partial_0 p_0=0 \  
\end{eqnarray}
and hence $p_0$ is constant. 

The equations of motion for $Y^K,K=p+1,\dots,9$ and $p_K$ have the form
\begin{eqnarray}\label{eqY}
& &	\partial_0 Y^K=\pb{Y^K,H}=2N \delta^{KL}p_L=-\frac{1}{p_0}\delta^{KL}p_L \ , \nonumber \\
& &\partial_0 p_K=\pb{p_K,H}=0 \ , \nonumber \\	
\end{eqnarray}
due to the fact that space-time metric does not depend on $Y^K$. Solving 
(\ref{eqY}) we get
\begin{equation}
	Y^K=-\frac{1}{2p_0}\delta^{KL}p_L^c\xi^0+Y^K_0 \ , \quad  p_K^c=\mathrm{const} \ 
\end{equation}
and this solution corresponds to the free motion of the the center of mass coordinates $Y^K$ in the transverse space.

Following \cite{Garousi:2004rd}  we write  the potential $V(T,X)$ as
\begin{equation}
V(T,l)=\mV(\rho)\sqrt{1+\frac{\rho^2}{l_s^2}l^Ml^N\eta_{MN}} \ ,
\end{equation}
where $\mV(\rho)$ has the same form as potential on the world-volume of unstable Dp-brane. 

In case of $\phi$ and $p_\phi$ we obtain
\begin{eqnarray}
\partial_\tau \phi=\frac{2N}{\rho^2}p_\phi \ , \quad 
\partial_\tau p_\phi=0 
\ . \nonumber \\
\end{eqnarray}
This equation shows that $p_\phi$ is constant that we choose to be equal to zero. Then $\phi=\mathrm{const}$ too.

Finally we consider equations of motion for $l^M$.
 We see that an interesting physics is hidden in the dynamics of $l^M$ and $\rho$ due to the fact
that the potential is function of these variables. In case of $l^M$ we obtain
\begin{eqnarray}
& &\partial_0 l^I=8N\delta^{IJ}p_J^l \ , \quad  
 \partial_0 p^l_I=
-8\frac{N}{l_s^2}\mV^2 \rho^2 \delta_{IJ}l^J
	\nonumber \\
\end{eqnarray}
that together give following differential equation for $l_I$
\begin{equation}
\partial_0^2 l_I=\frac{16}{p^2_0}\mV^2 \rho^2\delta_{IJ}l^J \ . 
\end{equation}
Further, the equation of motion for $\rho$ and $p_\rho$ have the from
\begin{eqnarray}
&&\partial_0 \rho=2N p_\rho \ , 
\nonumber \\
&&\partial_0 p_\rho=-8N\mV\frac{d\mV}{d\rho}(1+\frac{\rho^2}{l_s^2}
	l^Il_I)-8N\mV^2\frac{\rho}{l_s^2}l^Il_I \ . 
\end{eqnarray}
Clearly these equations of motion have solution when $l^I=0$ so that the tachyon obeys the equation
\begin{equation}
\partial^2_0 \rho=-\frac{4}{p_0^2}\mV\frac{d\mV}{d\rho} \ .
\end{equation}
that corresponds to standard rolling tachyon solution. 
Another interesting situation
occurs when we consider the tachyon at the minimum of the potential $\mV(\rho)$. Since, according to 
\cite{Garousi:2004rd} $\mV(\rho)$ has minimum at $\rho_{min}\pm \infty$ and $\mV(\rho)\sim e^{-\rho}$ for large $\rho$ we get that  $\lim_{\rho\rightarrow \rho_{min}}V \rho=0$.  
 Then  the equations of motion for $l^I$ have the form
\begin{equation}
\partial_0^2l_I=0 
\end{equation}
that has the solution $l_I=k_I \xi^0+l_I^{0}$ which of course is valid for  small values of $\xi^0$ since in the opposite case we would get out of the region of small  $l^I$ where our approximation is valid. In fact, more general analysis of the dynamics of D-brane anti D-brane system can be found in \cite{Garousi:2004rd}.

In summary, we performed canonical analysis of the Dp-brane anti-Dp-brane systems in the case of the vanishing gauge fields and in the approximation of the small separation of these objects. It would be certainly consider more general case and also it would be interesting to find covariant form of the action for Dp-brane anti Dp-brane system that was suggested in \cite{Israel:2011ut}. This problem is currently under investigation.

{\bf Acknowledgement:}
\\
This work 
is supported by the grant “Integrable Deformations”
(GA20-04800S) from the Czech Science Foundation
(GACR). 

\end{document}